\documentclass[twocolumn]{article}

\usepackage[utf8]{inputenc}
\usepackage[T1]{fontenc}
\usepackage{lmodern}
\usepackage{geometry}
\usepackage{graphicx}
\usepackage{amsmath, amssymb}
\usepackage{cite}
\usepackage{listings}
\usepackage{stfloats}
\usepackage{subcaption}
\usepackage{float}
\usepackage{siunitx}
\usepackage[table]{xcolor}
\usepackage{pgf}
\usepackage{hyperref}
\usepackage{xcolor}

\let\originalrmdefault\rmdefault
\let\originalsfdefault\sfdefault
\let\originalttdefault\ttdefault

\usepackage{bera}

\renewcommand{\rmdefault}{\originalrmdefault}
\renewcommand{\sfdefault}{\originalsfdefault}
\renewcommand{\ttdefault}{\originalttdefault}

\newcommand{\heat}[1]{%
\begingroup
\pgfmathparse{100*(#1/1.533)}%
\xdef\temp{\pgfmathresult}%
\endgroup
\cellcolor{red!\temp!white}#1
}

\colorlet{punct}{red!60!black}
\definecolor{background}{HTML}{EEEEEE}
\definecolor{delim}{RGB}{20,105,176}
\colorlet{numb}{magenta!60!black}

\lstdefinelanguage{json}{
    basicstyle=\fontfamily{fvm}\selectfont\footnotesize,
    numbers=left,
    numberstyle=\scriptsize,
    stepnumber=1,
    numbersep=8pt,
    showstringspaces=false,
    breaklines=true,
    frame=lines,
    backgroundcolor=\color{background},
    literate=
     *{0}{{{\color{numb}0}}}{1}
      {1}{{{\color{numb}1}}}{1}
      {2}{{{\color{numb}2}}}{1}
      {3}{{{\color{numb}3}}}{1}
      {4}{{{\color{numb}4}}}{1}
      {5}{{{\color{numb}5}}}{1}
      {6}{{{\color{numb}6}}}{1}
      {7}{{{\color{numb}7}}}{1}
      {8}{{{\color{numb}8}}}{1}
      {9}{{{\color{numb}9}}}{1}
      {:}{{{\color{punct}{:}}}}{1}
      {,}{{{\color{punct}{,}}}}{1}
      {\{}{{{\color{delim}{\{}}}}{1}
      {\}}{{{\color{delim}{\}}}}}{1}
      {[}{{{\color{delim}{[}}}}{1}
      {]}{{{\color{delim}{]}}}}{1},
}

\title{Turn-Based Combat Arena: A New Framework for Multiagent Training and Game Balancing }
\author{
V. M. Vasyuta$^{1}$
\and V. V. Malitskyi$^{2}$
\and O. S. Kushnir$^{1}$
\and B. I. Horon$^{1}$
\and V. A. Franiv$^{1}$
}
\date{}

\begin{document}

\twocolumn[
\maketitle
\begin{abstract}
This paper is the first in a series on Turn-Based Combat Arena, a configurable framework for turn-based strategy games designed to support the efficient training and evaluation of machine learning agents. The proposed framework enables flexible modification of game rules and parameters, allowing rapid experimentation across diverse scenarios. Its architecture is optimized for high-throughput simulation, supporting tens of thousands of games per second and enabling the storage and processing of billions of gameplay records on a single machine. The problem of balancing the game, and particularly the parameters of game units, is investigated in detail. We evaluate several optimization approaches and show that multiple methods converge to comparable solutions, suggesting robustness in identifying balanced game configurations. These results indicate that the framework can serve as a practical platform for both game design analysis and agent training.
\end{abstract}
\vspace{0.5cm}
]

\footnotetext[1]{%
Department of Optoelectronics and Information Technologies, Ivan Franko National University of Lviv, 107 Tarnavskyi Street, 79017 Lviv, Ukraine.
\texttt{waswasiuta@gmail.com},
\texttt{oleh.kushnir@lnu.edu.ua},
\texttt{bohdan.horon@lnu.edu.ua},
\texttt{volodymyr.franiv@lnu.edu.ua}
}

\footnotetext[2]{%
Gemicle, 24 Soborna Street, 21050 Vinnytsia, Ukraine.
\texttt{vadym.malitsky357@gmail.com}
}

\section{Introduction}
Since the early 1990s, when Chinook became the first computer program to defeat a human world champion in a board game (checkers) \cite{chinook1}, games have served as a crucial testbed for developing and evaluating artificial intelligence (AI) algorithms. They provide a controlled environment with clear rules and objectives, making the games ideal for testing various AI techniques.

Similar to Chinook, most AI players that first demonstrated superhuman performance in different games relied on databases of previously played games to determine the optimal behavior in the opening and endgame, while exploring intermediate game states, using some form of tree search. This is the case for chess (DeepBlue, 1997) \cite{chess-champion}, Othello (Logistello, 1997) \cite{logitello}, and Arimaa (SHARP, 2015) \cite{arimaa}. Such approaches require not only advanced algorithms, but also extensive computational resources. This can be seen in the comparison between game complexity and the year in which machine learning (ML) algorithms reached superiority in it. In particular, we deal with the problems with the state-space complexities of around $10^{20}$ for checkers \cite{checkers}, and $10^{43}$ for chess \cite{shannon-chess} and Arimaa \cite{arimaa_thesis}. As a result, mastering so complicated a game as Go with the state-space complexity $10^{170}$ \cite{go-complexity} or even more complex commercial video games has been a challenge for decades.

To overcome these challenges, in 2016, AlphaGo \cite{alpha-go}, an AI agent to play the game of Go, leveraged the power of neural networks and introduced a \textit{value network} for the estimation of state rewards and a \textit{policy network} for the selection of moves. This approach has significantly reduced the search space and allowed AlphaGo to achieve superhuman performance, despite the enormous search space.

AlphaGo's successor, AlphaZero \cite{alpha-zero}, has generalized this approach, demonstrating that the same architecture can achieve superhuman performance in chess and shogi without game-specific adjustments or a database of human-played games. The training process relies solely on self-play and reinforcement learning. A remarkable example can also demonstrate the power of neural networks: replacing a hand-crafted evaluation function in an open-source Stockfish ML player \cite{stockfish_github} with a fully neural network-based one \cite{stockfish_nn} has resulted in a significant performance improvement, allowing the player to defeat the long-time champion AlphaZero in 2025 \cite{stockfish_winner}.

Moreover, reinforcement learning models combined with neural networks have been successfully applied to commercial video games with practically infinite state spaces, such as StarCraft II (with approximately $10^{26}$ possible actions at each step) \cite{starcraft1}, Minecraft \cite{minecraft1}, Dota 2 \cite{dota2_1} (where OpenAI Five has defeated world champions in 2019) and Google Research Football \cite{google_football}.

Although neural networks, in conjunction with classical methods, currently represent the state of the art in gameplaying AI, achieving victory is not the only goal. There are many other open questions in game theory, such as the relation between player performance and complexity of his/her behavior and underlying decision-making processes \cite{chess_new}, the effects of diversity of playing strategies in multiagent gaming environments \cite{diversity1}, \cite{diversity2}, \cite{diversity3}, or the mechanisms of appearance of power laws and memory effects in game-history databases \cite{chess_zipf}, \cite{chess_memory} and in models' training and inference datasets \cite{alphazero_zipf}. Another open question of primary importance concerns the meaning of balance in the game \cite{balancing_wargames}. Finally, among all known board games, checkers remains the only one that has been (weakly) solved: it has been proven that the game always ends in a draw with perfect play from both sides \cite{checkers}.

These considerations motivate the investigation of simpler games than modern commercial ones, so that the basic principles of strategic behavior are not obscured by complicated game mechanics or the parameters of neural network agents. At the same time, the games under test must remain sufficiently complex to contain a diverse set of possible strategies. In this regard, turn-based strategy games are a promising domain for such studies.

From the game theory perspective, turn-based strategy games represent stochastic Markov games, first introduced by L. S. Shapley \cite{shapley} and actively studied in recent decades, with many important results obtained for equilibrium points \cite{mpe1,mpe2}, the value of games \cite{formula} and their computational complexity \cite{complexity}. Moreover, turn-based strategy games are widely represented in the gaming industry, making their study important from a practical standpoint.

There are many possible variations in the rules of turn-based strategy games \cite{tubstap}, but only a few have been widely considered in the literature. One of the most studied games of this type is TUBSTAP \cite{tubstap}, in which two teams composed of different types of units, with different attack, defense, attack-range, and movement-capacity parameters, aim to eliminate all units of their opponent on a two-dimensional rectangular board. Another game, STRATEGA \cite{stratega}, is an attempt to build a configurable game framework in which available units, map terrain, and even victory conditions can be configured via a YAML file. A more complicated game, ADAPTA \cite{adapta}, introduces a `factory' of game units, keeping the game itself relatively simple compared to commercial video games.

In this paper, we introduce a new, highly configurable game framework, the Turn-Based Combat Arena (TBCA), which is designed for ML studies. The game is described in Section \ref{the_game}, where we define its rules and derive an exact expression for its state-space complexity. In Section \ref{framework_architecture}, we describe the architecture of our framework, which allows one to achieve a game simulation throughput of tens of thousands of games per second and the storage and processing of billions of battle records on a single machine. Section \ref{game_balancing} is devoted to the central problem of this study, the balance of the game. Finally, in Section \ref{conclusion} we summarize the main results and conclusions.

\section{The Game}
\label{the_game}
\subsection{TBCA rule set}
\label{game_rules}

The idea of TBCA has been mainly inspired by the battle system of the classic \textit{Heroes of Might and Magic} series. However, it has been greatly simplified and reduced to a level resembling the battle systems of online browser games such as \textit{Forge of Empires}. Both games feature a battle of two teams consisting of units of different types, where each unit can attack a unit of the opposing team and reduce its health points. Such games are antagonistic and zero-sum, so that one team wins when it eliminates all of the opposing team's units.

Although TBCA shares some rules of gameplay with TUBSTAP, it differs from it in several important aspects. In particular, the movement order in TBCA is determined by the units' movement points, whereas players can move all units during their turn in TUBSTAP. The damage mechanics are also different: in TBCA, damage is defined by a configurable function and can include randomness. This makes TBCA more similar to the aforementioned commercial video games.

\begin{figure}[h]
\centering
\includegraphics[width=\linewidth]{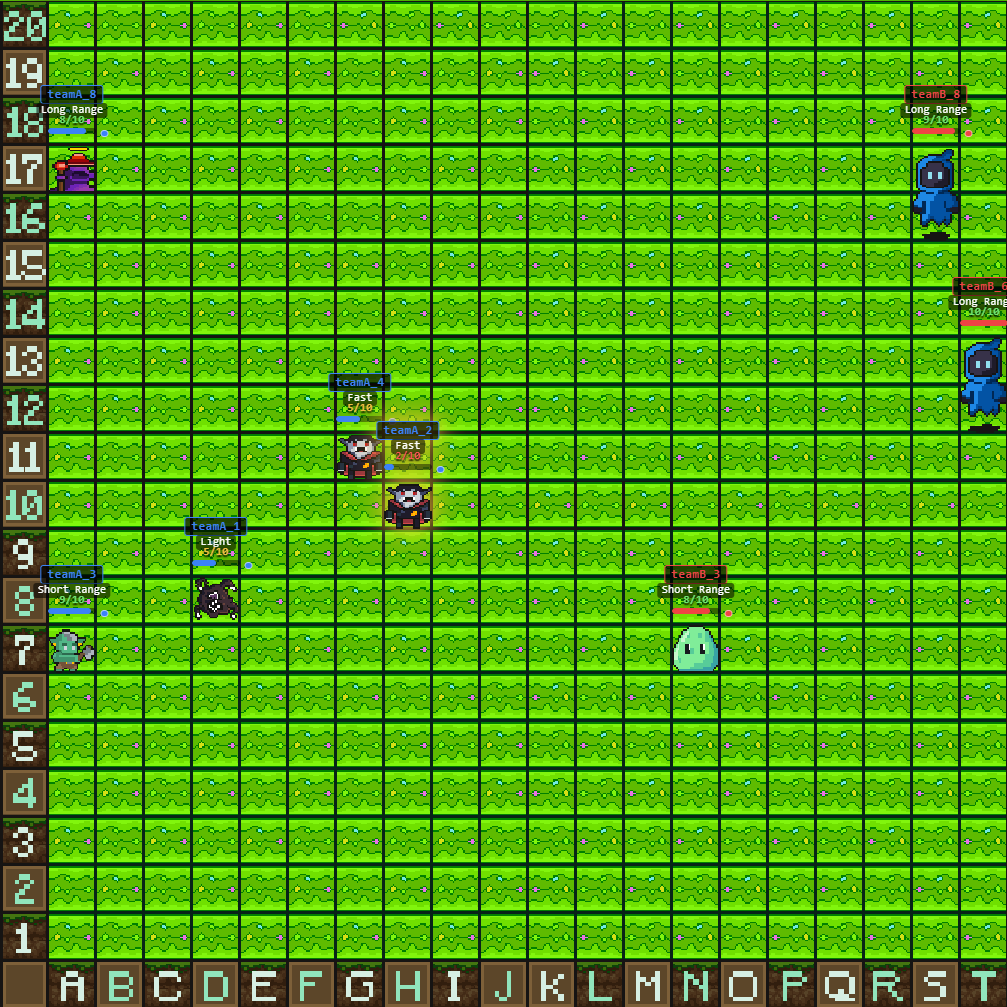}
\caption{A screenshot taken from the TBCA user interface. Team A, indicated by blue labels and health bars, has five remaining units: \textbf{L} at `D8', \textbf{F} at `H10' and `G11', \textbf{SR} at `A7', and \textbf{LR} at `A17'. Team B, indicated by red labels and health bars, has three remaining units: \textbf{SR} at `N7' and \textbf{LR} at `T12' and `S16'}
\label{fig:tbca}
\end{figure}

The game is highly configurable, and different configurations may lead to different optimal strategies. In the present paper, we consider a configuration that is closest to the \textit{Forge of Empires} game.

The TBCA game consists of elements described in the following Subsections.
\subsubsection*{Game geometry}
A battle is played in a rectangular $N\times M$ arena (the condition $N=M=20$ is specified in this study, although the dimensions of the arena, along with the other parameters, are configurable). The distances between orthogonally adjacent cells (top, bottom, left, and right) are equal to 1, while the distances between diagonally adjacent cells are set to 1.5.

Every cell has coordinates. Similarly to chess, each column of the arena is denoted by a Latin letter, and each row has a number, so cells' coordinates are represented as `D5', `H7', etc (see Figure~\ref{fig:tbca}).

A more complex topology of the map, with barriers and cells that confer bonuses or impose taxes, is also possible but is not considered in the present paper.

\subsubsection*{Players}
There are two teams, \textit{A} and \textit{B}, each containing $T$ units ($T=8$ in this study). At the beginning of the game, units are placed in the leftmost (`A') and rightmost (`T') columns, respectively. In the configuration considered below, units are spawned at deterministic positions corresponding to the odd numbers from 3 to 17. However, they can also be spawned at random positions.

\subsubsection*{Units}
\label{units}
A unit that participates in a game can belong to one of five configurable types. Each of these types, in fact, represents an arbitrary class of elements (`soldiers'). For the sake of clarity, we adopt the terminology used in \textit{Forge of Empires}: Light (\textbf{L}), Heavy (\textbf{H}), Fast (\textbf{F}), Short-Range (\textbf{SR}), and Long-Range (\textbf{LR}). In the context of ancient or medieval warfare, Light units \textbf{L} can be viewed as light infantry or melee fighters equipped with light armor and weapons, making them more mobile than heavily armored \textbf{H} infantry, at the expense of lower defensive capabilities. Fast units \textbf{F} can serve as a representation of cavalry, which is much faster than infantry. Short-Range units \textbf{SR} can be viewed as slingers or archers capable of attacking from a relatively short distance, while Long-Range units \textbf{LR} can represent ballistae or other artillery with a much longer attack range than regular archers \textbf{SR}. Each unit type has a unique (configurable) set of parameters:

\begin{itemize}
\item \textit{Health level}: a natural number in the range $0-H$ (to be specific, we adopt $H=10$). A unit dies when its health level reaches zero. Different units can differ in their initial health levels, but here we use the same initial health level for all units;
\item \textit{Attack level} $A$: a natural number representing the attack modifier. We stress that this is a modifier rather than an absolute attack value, and therefore $A=0$ does not necessarily mean that a unit cannot attack. Rather, its attack power will be minimal. The exact effect of $A$ is defined by the damage function given by formula (\ref{damage_function}) below;
\item \textit{Defense level} $D$: a natural number representing the defense modifier. As with $A$, this is a modifier rather than an absolute value, and everything said about $A$ can also be mirrored for $D$;
\item \textit{Attack range} $AR$: a natural number. From its current position, a unit can attack an enemy without moving if the integer part of the distance between the cells occupied by the two units is less than or equal to $AR$. For example, $AR=1$ means that a unit can attack enemies in all neighboring cells, including diagonally adjacent cells;
\item \textit{Movement points} $MP$: a natural number. It is responsible for both the movement order and the maximum distance a unit can move during a single movement. A unit can move from one cell to another only if the integer part of the distance between them is less than or equal to $MP$. Two units cannot occupy the same cell, but a unit can move through an occupied cell without penalty.
\end{itemize}

A health level $H$ provides a unit with the ability to survive several attacks. The attack range $AR$, and the movement points $MP$ define the tactical properties of the units, e.g., melee fighters \textbf{L} and \textbf{H} or archers \textbf{SR}. The attack ($A$) and defense ($D$) levels serve as balancing parameters (see the following Sections): their purpose is to compensate for possible tactical advantages associated with a specific attack range and the number of movement points.

The type of unit in a particular game can be defined deterministically or randomly, thus enriching the game with different modes. For example, the aim of the game might be to win a series of consecutive battles against predefined teams of units, with each unit's type known in advance. In this case, the game is more deterministic and allows for a more strategic approach. Alternatively, the type of each unit in a team can be randomly determined at the start of each battle, making the game more stochastic and requiring players to adapt their strategies on the fly. Since our main interest is to study the statistical properties of AI agents' behavior and their ability to adjust to a changing environment, we primarily consider the latter case in the present study, in which the type of each unit is randomly determined at the beginning of each battle.

\subsubsection*{Damage mechanics}
When a unit attacks an enemy, the target loses a certain number of health points. If the target survives, it can retaliate whenever the attacker is within its attack range. In particular, if the attacker is a melee unit, retaliation is inevitable, whereas if a longer-range unit attacks a melee unit, the latter cannot retaliate. According to the rules of the game, the health points lost during retaliation are half those lost from a direct attack.

A damage function determines the harm inflicted on a unit by an attack. The damage function $DM$ should depend on the attack level $A$ of the attacker and the defense level $D$ of the target. Without loss of generality, we represent $DM$ as follows:
\begin{equation}
\label{damage_function}
DM = DM_0 \cdot \mathcal{A}(A,D) \cdot \mathcal{R},
\end{equation}
where $DM_0$ is the basic damage, which determines the average number of attacks $\mathcal{N}$ that a unit can survive. It is therefore reasonable to define it as $DM_0 = H/\mathcal{N}$. Finally, $\mathcal{A}(A, D)$ is a modifier that accounts for the impact of the attacker's level of attack $A$ and the target's level of defense $D$. Here, $\mathcal{R}=1+\alpha \operatorname{rnd}(-1,1)$ denotes a random factor modifier, $\alpha$ a randomness amplitude, and $\operatorname{rnd}(-1,1)$ a random variable distributed with some probability in the range from $-1$ to 1.

The damage function should return an integer value. In general, it depends on one's definition of how to deal with the appropriate fractional part. However, to cover the cases where each attack should result in a health loss, it makes sense to define it as $DM = \lfloor DM_0\cdot\mathcal{A}(A, D)\cdot\mathcal{R}\rfloor$, where $\lfloor ... \rfloor$ means the floor function.

\subsubsection*{Movement decider}
In board games such as chess and turn-based strategy games such as TUBSTAP, players move their units alternately. In TBCA, as in many other commercial video games of this kind, the movement priority is determined by a unit's movement points. At the beginning of the game, each unit on the board is assigned a rank based on its movement points. The rank remains constant throughout the game. A higher number of movement points results in an earlier turn. The movement order of units with the same number of movement points is randomized. This randomization prevents any bias related to the selection of either the `left-hand' or `right-hand' team.

During each turn, a unit can perform one of several actions: skip its turn (i.e., do nothing), move to an available cell, attack an opponent without moving, or move and then attack. Multiple movements or attacks during a single turn are forbidden. A unit also cannot attack and move after that attack during the same turn.

\subsubsection*{Draws}
Under some circumstances, a game can run indefinitely long without a winner. For example, this can occur if both players repeatedly make the same moves without attacking their opponent. This situation can become an obstacle for ML model training, especially on maps that are large relative to the units' movement ranges. To avoid this situation, a draw is declared if the number of `idle' turns exceeds a predefined configurable threshold.

\subsection{State-space complexity}
In TBCA, each team consists of $T$ units. Each unit in a team can be one of five types and can have one of $H$ possible health levels (we assume that all units have the same initial health level). The absence of a unit is formally expressed by its zero health value (i.e., the unit is dead). Using the stars-and-bars combinatorial formula, one can calculate the number $N_1$ of different states of all units in a team as follows:
\begin{equation*}
N_{1} = \binom{T + 5(H+1)-1}{5(H+1)-1}=\binom{62}{54}\sim3.38\cdot10^9.
\end{equation*}
Since there are two teams in a game, the factor that takes into account the number of different states of both teams is equal to $N_1^2$. Finally, taking into account the spatial arrangement of units and the constraint that each cell can be occupied by at most one unit, we obtain the following expression for the state-space complexity of the game:
\begin{equation*}
\mathcal{C}=\binom{T + 5(H+1)-1}{5(H+1)-1}^2\frac{(NM)!}{(NM-2T)!}\sim3.62\cdot10^{60},
\end{equation*}
which, for the chosen parameters, exceeds the state-space complexity of chess ($10^{43}$) and lies between the corresponding values of Hex ($10^{57}$) and Thurn and Taxis ($10^{66}$) \cite{thurn}.

By modifying the maximum health level, the number of unit types, and the number of units per team, the state-space complexity of the game can be adjusted, making the game more or less complex. This establishes a connection between simple games with low state-space complexity, where an optimal strategy can be found using brute-force search, and complex games with high state-space complexity, where the optimal strategy remains unknown and can only be approximated using advanced ML methods.

\section{Framework Architecture}
\label{framework_architecture}
Training any ML model involves generating extensive training datasets. This imposes additional requirements on the architecture of our framework, demanding the following: (a) a high game simulation throughput, (b) effective metrics collection and rapid incorporation of new data into the model during reinforcement learning, and (c) an easy way to integrate new ML models.
\begin{figure}[h]
\centering
\includegraphics[width=\linewidth]{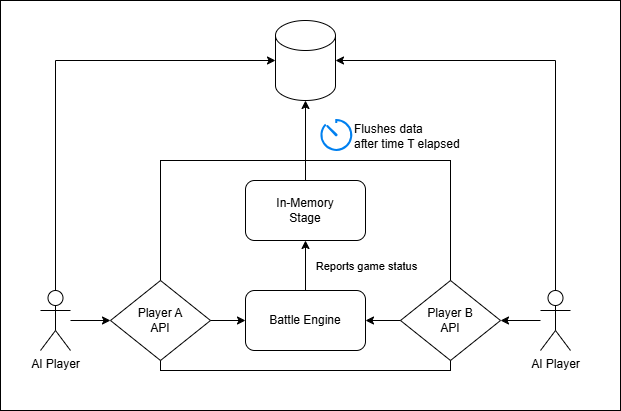}
\caption{Diagram explaining the architecture of our framework.}
\label{fig:architecture}
\end{figure}

To fulfill the above-mentioned requirements, we have implemented our system with the following components (see Figure~\ref{fig:architecture}):
\begin{itemize}
\item \textit{Battle engine}. This module implements the battle logic and game rules described in Subsection \ref{game_rules}. The engine is stateless by design, allowing one to run it simultaneously in many parallel threads. As a result, when benchmarking random battles within the TBCA framework, we have achieved a game rate of 55,000 games/s. Given that units traverse approximately 120 states per game on average, this corresponds to a state generation rate of approximately $6\cdot10^6$ states/s. Note that the appropriate tests have been conducted with the engine running inside a Docker container with resource limits of 16 CPU cores and 16 GB of RAM. All benchmarks have been performed on a single machine equipped with an Intel Core i5-13500K processor and 32 GB of RAM;
\item \textit{Database}. Given the number of parameters that can describe a battle state, the number of possible features can theoretically be as large as the state-space complexity of the game. In practice, however, it is necessary to reduce the dimensionality of the problem by extracting only the most relevant features. Nevertheless, the system should be designed to handle as large a dataset as possible. Modern OLAP column-oriented storage systems are well-suited for processing large amounts of such data. For example, the local DuckDB database used in this work can efficiently handle 300--400 million records while occupying approximately 100 GB of local disk space;
\item \textit{In-memory stage}. A drawback of using column-oriented storage systems for large-scale data processing is their limited ability to handle frequent updates (typically not more than a few updates per second). To address this limitation, we have introduced an in-memory stage that accumulates the required battle statistics and periodically flushes them into the database according to a predefined schedule;
\item \textit{Battle API}. To facilitate the integration of arbitrary AI players into the system, we have developed a real-time WebSocket API. Whenever a player has to act, it receives information about the current battle state and the next acting unit as follows:
\begin{lstlisting}[language=json,firstnumber=1]
{
    "battleID": "nJRzakFiTB1NN",
    "nextUnitInfo": {
        "teamName": "teamA",
        "unit": {...},
        "availableDestinations": [...],
        "availableTargets": [...]
    },
    "teamA": {
        "name": "teamA",
        "units": [...]
    },
    "teamB": {...}
}
\end{lstlisting}

In response, the player must report one of the possible actions: `Skip', `Attack', or `Move', along with the corresponding parameters and the optional label:
\begin{lstlisting}[language=json,firstnumber=1]
{
    "actionType": "Skip" | "Attack" | "Move",
    "destination": "...",
    "target": "...",
    "label": "..."
}
\end{lstlisting}

The field `label' allows an AI player to attach metadata to each decision. For example, a model can operate at the level of high-level strategies, with the label indicating which strategy has been selected at a particular step. This information can be useful for model tuning and optimization. It is stored in the database alongside other battle statistics, thus allowing subsequent analysis of the model's behavior. The label is optional and can be omitted if the agent does not use it.
\end{itemize}

The system also includes several built-in rule-based combat players that follow different strategies. These players can use the same Battle API as external AI players, making them interchangeable with externally developed agents.

\section{Game Balancing}
\label{game_balancing}

Balancing the game is essential to ensure fairness and prevent the emergence of dominant units or plainly dominating strategies. From a human player's perspective, such dominance would have made it less appealing to use weaker units or follow inferior strategies, reducing the set of viable strategies and making the game less engaging. In a game mode where teams are randomly selected, such an imbalance can be mitigated by a sufficiently large number of games, as stronger units are expected to appear equally often in both teams over many rounds. However, this does not fully resolve the issue. In practice, imbalances can still distort battle outcomes, making them overly dependent on the presence of particular units rather than on player strategy. This not only diminishes the game's strategic depth but can also frustrate players. Moreover, this imbalance can introduce systematic biases into the simulation data, thus complicating the training of ML models.

Although the problem of game balance has been repeatedly studied from different perspectives \cite{balance1, balance2, balance3,liitman,balancing_wargames, balance00}, there is still no generic definition of game balance or a generally accepted approach to defining a loss function for the corresponding optimization problem. One possible approach to game balance is to compare different strategies and ensure that no single strategy dominates all others \cite{balance1, balance2, balance3}. During consecutive simulations, AI players may adopt weights over different strategies from a predefined set. However, the number of possible strategies can be very large; the optimal strategy may itself be a combination of multiple strategies, and it can even depend on the specific strategy employed by the opponent \cite{liitman}. This makes such an approach difficult to formalize rigorously.

Another viewpoint is to balance the parameters of units. Classical approaches to unit parameter balancing include analyzing pairwise win-loss matrices from a sufficiently large number of games between teams consisting of identical units \cite{balancing_wargames, balance00}, to achieve equal average win rates across all unit types.

In this paper, we focus on balancing the parameters of game units. However, requiring equal win rates across all unit types is not the only possible criterion. In this Section, we introduce different metrics for unit parameter balancing, treat them as loss functions for the corresponding optimization problem, and compare the optimization results obtained by these metrics.

\subsection{Balance metrics}
\subsubsection*{Pairwise win-loss matrix for the same-unit teams}
A possible set of metrics comes from the idea of comparing the effectiveness of a team composed of units of one type against that of a team composed of units of another type. One can simulate many battles for all possible pairs of unit types, compute the average win rate for each type, and attempt to equalize these rates across all unit types.

However, the win rate is too coarse and does not reflect the strength of units with sufficient granularity. For example, there is a substantial difference between a situation in which an \textbf{LR} team defeats an \textbf{H} team before the latter reaches the artillery units \textbf{LR}, on the one hand, and a situation in which the \textbf{H} units manage to engage the enemy and eliminate half of the opposing team, on the other hand. From this perspective, counting the number of enemies killed, or even the total damage dealt, provides a more precise metric.

Let us denote by $\overline{k^i_j}$ the average number of kills per battle achieved by a team composed of units of type $i$ when playing against a team composed of units of type $j$, where $i,j = 1,\ldots,5$, $i \neq j$, and where the unit types are defined as in Subsection \ref{units}. Here, the overline operator $\overline{x}$ denotes the averaging over simulations. Together, the values $\overline{k^i_j}$ form a hollow, non-symmetric matrix $5\times5$. The values of $\overline{k^i_j}$ range from $0$, when the $i$-team is unable to eliminate any opponent unit in any game, to $T$, when it wins all battles without losing a single unit.

The balance can be thought of as the situation in which the average kill rate of each type $i$ of units against all the other types of units $j$ tends to a kill rate value $\kappa$ averaged over teams: $\langle k\rangle^i=\frac{1}{4}\sum_{j\neq i}\overline{k^i_j}\rightarrow \kappa$, where $\kappa=\frac{1}{5}\sum_i\langle k\rangle^i$. The same condition can be imposed on the average death rate of units of type $j$, which should also converge to the same value $\kappa$: $\langle k\rangle_j=\frac{1}{4}\sum_{i\neq j}\overline{k^i_j}\rightarrow \kappa$. Indeed, we have
\begin{align*}
\kappa &= \frac{1}{5}\sum_i\langle k\rangle^i
= \frac{1}{5\cdot4}\sum_i\sum_{j\neq i}\overline{k^i_j}
= \frac{1}{5\cdot4}\sum_j\sum_{i\neq j}\overline{k^i_j} \\
&= \frac{1}{5}\sum_j\langle k\rangle_j.
\end{align*}
Then we can define the normalized mean squared deviation $k^2$ from this balance as
\begin{eqnarray}
k^2 = \frac{1}{5}\sum_i\frac{1}{2}\left(\frac{\left[\langle k\rangle^i-
\kappa\right]^2}{\kappa^2} + \frac{\left[\langle k\rangle_i-
\kappa\right]^2}{\kappa^2}\right).
\end{eqnarray}

Moreover, nothing stops us from demanding a strict balance of kill-to-death rate $\overline{k^i_j}\rightarrow\kappa$, leading to the metric
\begin{eqnarray}
k^2_0 = \frac{1}{20\kappa^2}\sum_{i,j,i\neq j}\left(\overline{k^i_j}-\kappa\right)^2.
\end{eqnarray}

In the same manner, one can consider the average pairwise damage $\overline{d^i_j}$ instead of the number of kills, resulting in the metrics
\begin{eqnarray}
d^2 = \frac{1}{5}\sum_i\frac{1}{2\delta^2}\left(\left[\langle d\rangle^i-
\delta\right]^2 + \left[\langle d\rangle_i-
\delta\right]^2\right),
\end{eqnarray}
where $\langle d\rangle^i=\frac{1}{4}\sum_{j\neq i}\overline{d^i_j}$, $\langle d\rangle_j=\frac{1}{4}\sum_{i\neq j}\overline{d^i_j}$, $\delta=\frac{1}{20}\sum_{i,j,i\neq j}\overline{d^i_j}$, and the metrics
\begin{eqnarray}
d^2_0 = \frac{1}{20\delta^2}\sum_{i,j,i\neq j}\left(\overline{d^i_j}-\delta\right)^2.
\end{eqnarray}

Although $d^2$ is likely a more precise metric than $k^2$, it suffers from great variance. In fact, a sample of 10 evaluations of $k^2$-type metrics has yielded a measurement error of up to 0.3\%, with 2,000,000 games played per evaluation. In contrast, the same number of evaluations for $d^2$-based metrics results in an error of up to 14\%, even with 5,000,000 simulations per evaluation. This high level of fluctuation makes $d^2$-like metrics less practical as loss functions for the unit parameter balancing optimization problem. However, our simulations have shown that $k^2$-like and $d^2$-like metrics are correlated: larger values of $k^2$ generally correspond to larger values of $d^2$, and vice versa (see Table~\ref{tab_results}).

\subsubsection*{Pairwise win-loss matrix for semi-random teams}
Comparing teams composed exclusively of a single unit type is not always realistic, as game scenarios often involve random team compositions. In a randomly composed team, the expected number of units $n_i$ of a specific type $i$ in such a team is given by $\mathbb{E}[n_i] = T/5 = 1.6$. Moreover, teams consisting of units of a single type do not capture interactions between different unit types. For example, if a team consists of both \textbf{H} and \textbf{LR} units, the latter can attack from a distance, while the Heavy units \textbf{H} absorb most of the incoming damage. In such a scenario, it may be beneficial to increase the defense level $D$ of the \textbf{H} units, improving their ability to withstand attacks and giving the \textbf{LR} units of the same team more time to deal damage without retaliation.

Therefore, we should also consider simulations of team battles in which some units are randomly selected, while others belong to a specific type. In our setting, we construct teams containing five units, one of each type, while the remaining $(T-5)$ units are of a selected type. This leads to the metrics $k'$, $d'$, $k'_0$, and $d'_0$, which are analogous to those introduced by formulae (2)--(5), but are defined for the battles involving semi-random teams.
\subsubsection*{Battle-level metrics}
The metrics defined above have been introduced under the assumption that each unit's contribution is identical across teams, regardless of its teammates. Unfortunately, these metrics do not comprehensively account for random team compositions, which are the main focus of the game mode under investigation. We therefore introduce additional metrics specifically designed for battles with randomly composed teams.

For example, the fairness of the game can be reasonably associated with a tendency toward draws. In a well-balanced game, the number of units survived in the winning team should be small, since units are expected to be expended in the process of eliminating opponents of statistically similar strength. Based on this intuition, it is natural to consider the mean squared error (MSE) $\sigma_n$ of the number $n$ of units remaining at the end of a game. Here, we take $n>0$ if the `left-hand' player wins and $n<0$ otherwise. Since the movement turn depends only on the movement points of the units and not on their side on the board, we arrive at the condition $\mathbb{E}[n]=0$.

Additionally, a well-balanced game should exhibit lower variability in the total number $a$ of actions than an imbalanced game. For this reason, we consider the MSE $\sigma_a$ of the total number of actions, which is normalized by the number $\overline{a}$ of actions averaged over all games, i.e., the parameter $\varepsilon_a=\sigma_a/\overline{a}$.

These metrics do not necessarily represent the best candidates for loss functions, at least because there can be limiting cases where they take small values, even though the game remains highly imbalanced. For example, our experience is that the battles can become short and thus exhibit little variation $\varepsilon_a$ when \textbf{SR} and \textbf{LR} units are characterized by large values of different parameters, while all other units have small parameter values. This would result in small values of both $\sigma_n$ and $\varepsilon_a$. Nevertheless, we will later show that these metrics remain useful for analyzing optimization results using other loss functions.

\subsubsection*{Unit type-specific metrics}
One may also be interested in the metrics that describe the performance of a specific unit type $i$. An example is the survival rate percentage $s_i=\overline{S_{i}}/\overline{N_{i}}$, where $\overline{S_{i}}$ is the average number of units of a type $i$ that remain alive at the end of battle, and $\overline{N_{i}}$ is the initial number of units of that type averaged over all games played. For fully random team compositions, we have $\overline{N_{i}}=T/5$. One can consider the MSE $\sigma_s$ of the survival rates found across all types of units as a loss function for the corresponding optimization problem: $\sigma_s = \frac{1}{5}\sum_i(s_i-s)^2$, where $s=\frac{1}{5}\sum_i s_i$.

We define another metric, the victory impact $w_{i}$, as the average $w_i=\overline{W_{i}}$, where $W_{i}$ is the initial number of units of type $i$ in the winning team. If the game is fairly balanced and all units are expected to be equally effective, one can demand that, for randomly composed teams, the condition given by $w_i\rightarrow T/5$ holds for all $i$. The MSE $\sigma_w$ of $w_i$ given by $\sigma_w = \frac{1}{5}\sum_i(w_i-w)^2$ (with $w=\frac{1}{5}\sum_i w_i$) can also be used as a candidate loss function.

\begin{figure*}[!t]
\begin{subfigure}{0.95\textwidth}
\centering
\includegraphics[width=\linewidth]{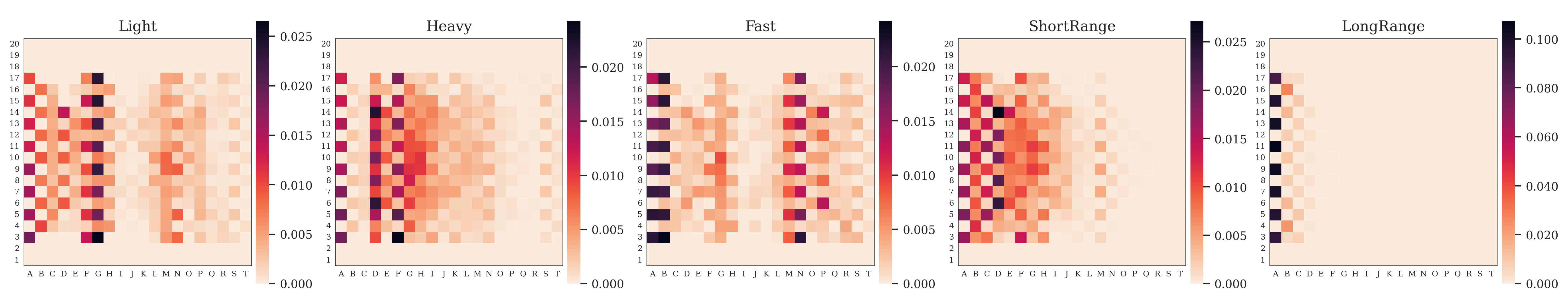}
\caption{Moves heatmap}
\end{subfigure}
\hfill
\begin{subfigure}{0.95\textwidth}
\centering
\includegraphics[width=\linewidth]{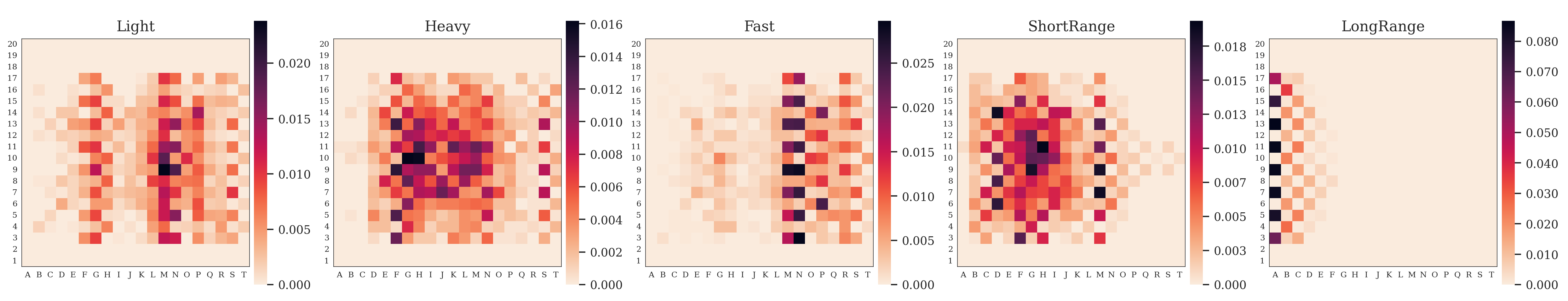}
\caption{Deaths heatmap}
\end{subfigure}
\caption{Heatmaps of unit-specific moves and deaths. Unit-specific (`left-hand') teams have played 1000 games with random-unit (`right-hand') teams. The parameters of units have been defined by $\sigma_w$-based optimization.}
\label{heatmaps}
\end{figure*}

\subsection{Pre-balancing setup and methodology}

\begin{table*}[!t]
\centering
\begin{tabular}{c|c|c||c|c|c|c|c|c}
\hline
Loss function & & $A$, $D$ & $s_i$ & $w_i$ & $k^2$ & $d^2$ & $\sigma_n$ & $\overline{a}, \varepsilon_a$ \\
\hline\hline
$k^2$&
$\begin{array}{c}\textbf{L} \\ \textbf{H} \\ \textbf{F} \\ \textbf{SR} \\ \textbf{LR}\end{array}$ &
$\begin{array}{c}15,19 \\ 30,15 \\ 25,8 \\ 30,8 \\ 13,0\end{array}$ &
$\begin{array}{c}3.6\% \\ 6.5\% \\ 3.4\% \\ 23.0\% \\ 48.2\% \end{array}$ &
$\begin{array}{c}1.60 \\ 1.48 \\ 1.32 \\ 1.75 \\ 1.84 \end{array}$ &
$\begin{array}{c}k^2: 0.024 \\ k_0^2: 0.533 \\ {k^\prime}^2: 0.119 \\ {k_0^\prime}^2: 0.160 \end{array}$ &
$\begin{array}{c}d^2: 0.064\pm0.006 \\ d_0^2: 0.229\pm0.016 \\ {d^\prime}^2: 0.088\pm0.014 \\ {d_0^\prime}^2: 0.138\pm0.013 \end{array}$ &
2.99 & $\begin{array}{c}199, \\ 0.068 \end{array}$ \\
\hline

$k_0^2$&
$\begin{array}{c}\textbf{L} \\ \textbf{H} \\ \textbf{F} \\ \textbf{SR} \\ \textbf{LR}\end{array}$ &
$\begin{array}{c}29,12 \\ 29,18 \\ 27,11 \\ 29,18 \\ 10, 0\end{array}$ &
$\begin{array}{c}4.0\% \\ 6.2\% \\ 4.2\% \\ 28.0\% \\ 44.0\% \end{array}$ &
$\begin{array}{c}1.63 \\ 1.43 \\ 1.38 \\ 1.87 \\ 1.69 \end{array}$ &
$\begin{array}{c}k^2: 0.297 \\ k_0^2: 0.462 \\ {k^\prime}^2: 0.108 \\ {k_0^\prime}^2: 0.148 \end{array}$ &
$\begin{array}{c}d^2: 0.125\pm0.008 \\ d_0^2: 0.236\pm0.010 \\ {d^\prime}^2: 0.079\pm0.011 \\ {d_0^\prime}^2: 0.136\pm0.016 \end{array}$ &
3.06 & $\begin{array}{c}199, \\ 0.059 \end{array}$ \\
\hline

${k^\prime}^2$&
$\begin{array}{c}\textbf{L} \\ \textbf{H} \\ \textbf{F} \\ \textbf{SR} \\ \textbf{LR}\end{array}$ &
$\begin{array}{c}27,0 \\ 8,18 \\ 24,5 \\ 16,1 \\ 0,5\end{array}$ &
$\begin{array}{c}3.8\% \\ 12.2\% \\ 5.2\% \\ 22.0\% \\ 41.4\% \end{array}$ &
$\begin{array}{c}1.60 \\ 1.61 \\ 1.56 \\ 1.62 \\ 1.61 \end{array}$ &
$\begin{array}{c}k^2: 0.454 \\ k_0^2: 0.707 \\ {k^\prime}^2: 0.006 \\ {k_0^\prime}^2: 0.045 \end{array}$ &
$\begin{array}{c}d^2: 0.134\pm0.004 \\ d_0^2: 0.228\pm0.007 \\ {d^\prime}^2: 0.039\pm0.010 \\ {d_0^\prime}^2: 0.091\pm0.012 \end{array}$ &
3.00 & $\begin{array}{c}217, \\ 0.089 \end{array}$ \\
\hline

${k_0^\prime}^2$&
$\begin{array}{c}\textbf{L} \\ \textbf{H} \\ \textbf{F} \\ \textbf{SR} \\ \textbf{LR}\end{array}$ &
$\begin{array}{c}28,11 \\ 7,30 \\ 29,11 \\ 17,5 \\ 0,10\end{array}$ &
$\begin{array}{c}4.0\% \\ 13.5\% \\ 5.1\% \\ 23.3\% \\ 41.6\% \end{array}$ &
$\begin{array}{c}1.60 \\ 1.66 \\ 1.52 \\ 1.62 \\ 1.61 \end{array}$ &
$\begin{array}{c}k^2: 0.455 \\ k_0^2: 0.693 \\ {k^\prime}^2: 0.011\\ {k_0^\prime}^2: 0.044 \end{array}$ &
$\begin{array}{c}d^2: 0.128\pm0.003 \\ d_0^2: 0.209\pm0.005 \\ {d^\prime}^2: 0.045\pm0.011 \\ {d_0^\prime}^2: 0.089\pm0.012 \end{array}$ &
3.10 & $\begin{array}{c}240, \\ 0.112 \end{array}$ \\
\hline

$\sigma_s$ &
$\begin{array}{c}\textbf{L} \\ \textbf{H} \\ \textbf{F} \\ \textbf{SR} \\ \textbf{LR}\end{array}$ &
$\begin{array}{c}29,29 \\ 8,29 \\ 29,27 \\ 7,0 \\ 0,1\end{array}$ &
$\begin{array}{c}29.7\% \\ 27.6\% \\ 30.6\% \\ 29.2\% \\ 39.9\% \end{array}$ &
$\begin{array}{c}1.98 \\ 1.42 \\ 1.94 \\ 1.28 \\ 1.37 \end{array}$ &
$\begin{array}{c}k^2: 0.640 \\ k_0^2: 0.891 \\ {k^\prime}^2: 0.383 \\ {k_0^\prime}^2: 0.506 \end{array}$ &
$\begin{array}{c}d^2: 0.396\pm0.004 \\ d_0^2: 0.550\pm0.003 \\ {d^\prime}^2: 0.251\pm0.011 \\ {d_0^\prime}^2: 0.339\pm0.011 \end{array}$ &
5.34 & $\begin{array}{c}250, \\ 0.127 \end{array}$ \\
\hline

$\sigma_w$ &
$\begin{array}{c}\textbf{L} \\ \textbf{H} \\ \textbf{F} \\ \textbf{SR} \\ \textbf{LR}\end{array}$ &
$\begin{array}{c}22,21 \\ 12,29 \\ 30,16 \\ 30,1 \\ 9,11\end{array}$ &
$\begin{array}{c}4.3\% \\ 13.4\% \\ 6.3\% \\ 22.1\% \\ 41.4\% \end{array}$ &
$\begin{array}{c}1.61 \\ 1.59 \\ 1.60 \\ 1.60 \\ 1.61 \end{array}$ &
$\begin{array}{c}k^2: 0.495 \\ k_0^2: 0.775 \\ {k^\prime}^2: 0.024 \\ {k_0^\prime}^2: 0.069 \end{array}$ &
$\begin{array}{c}d^2: 0.145\pm0.004 \\ d_0^2: 0.282\pm0.008 \\ {d^\prime}^2: 0.054\pm0.009 \\ {d_0^\prime}^2: 0.104\pm0.019 \end{array}$ &
3.12 & $\begin{array}{c}230, \\ 0.104 \end{array}$ \\
\hline

\end{tabular}
\caption{Balancing results obtained for different loss functions. Minimization of different loss functions is performed using the CMA-ES algorithm, with 10,000 simulations per candidate solution. Metrics $s_i$, $w_i$, $\sigma_n$, $\overline{a}$, and $\varepsilon_a$ are calculated using 100,000 simulations. Variants of metrics $k^2$ and $d^2$ are calculated using 100,000 and 250,000 simulations for each unit-type pair, respectively (i.e., 2,000,000 and 5,000,000 simulations in total). Values presented in the table are averages over 10 measurements. The measurement error does not exceed 0.3\% for all metrics, except for variants of $d^2$, and is equal to up to 25\% for $d^2$-like metrics. The confidence probability for all error calculations is 0.95. The errors are reported only for $d^2$-like metrics, since they remain at or below the level of numerical precision for all other metrics.
}
\label{tab_results}
\end{table*}

In principle, the game has many parameters that can be tuned to improve the balance. To be specific and avoid getting lost in the details, we focus exclusively on variations in the attack ($A$) and defense ($D$) levels and their impact on the game.

Then the tactical points ($AR$, $MP$) have been fixed at (1, 12), (1, 5), (1, 17), (7, 5), and (17, 2) respectively for Light (\textbf{L}), Heavy (\textbf{H}), Fast (\textbf{F}), Short-Range (\textbf{SR}), and Long-Range (\textbf{LR}) units. The rationale behind this setup is intuitive: far-reaching artillery \textbf{LR} covers the largest part of the arena with $AR_{LR}=17$, but remains very slow. The cavalry \textbf{F} unit is fast, with $MP_F = AR_{LR}$, ensuring that \textbf{LR} cannot attack it safely, without the risk of being attacked in return in the next round. However, \textbf{F} cannot immediately attack an enemy at the beginning of the game. Archers (\textbf{SR}) cannot reach so far as artillery \textbf{LR}, and we arrive at the condition $AR_{SR} + MP_{SR} = MP_{L}$ for the same reason as for \textbf{F} and \textbf{LR}. Finally, Heavy units \textbf{H} are slow, with $MP_H = MP_{SR}$.

The damage function (\ref{damage_function}) has been chosen based on the following considerations. We define the basic damage value as $DM_0 = H/3.5$. The denominator is slightly smaller than the linear size of the arena divided by the movement points of the slowest melee unit \textbf{H}, giving such units a higher chance of reaching artillery \textbf{LR} units before being eliminated. The power modifiers $A$ and $D$ are restricted to the range $0 \leq A, D \leq 30$, and we define $\mathcal{A}(A, D) = 1 + (A-D)/50$, which provides a sufficient range of influence on the power of a unit, with $0.4 \leq \mathcal{A} \leq 1.6$. Finally, the random modulation amplitude is fixed at $\alpha = 0.1$, and $\operatorname{rnd}(-1,1)$ is a uniformly distributed random variable, resulting in $0.9 \leq \mathcal{R} \leq 1.1$.

The process of balancing the game also depends on the agent(s) used for tuning. For this purpose, we have developed a rule-based combat agent that follows a simple algorithm: if a unit can attack someone, it attacks the closest enemy; otherwise, it moves to the closest position from which it can attack in the next round. Despite its simplicity, the proposed rule-based combat agent exhibits substantial behavioral diversity among different types of units. This can be seen in Figure~\ref{heatmaps}, where unit-specific movement and death heatmaps are shown.

In the end, the balancing problem is formulated as an optimization problem with $5 \cdot 2 = 10$ optimization parameters, each taking 31 possible integer values. We have considered different metrics such as $k^2$, $k_0^2$, ${k'}^2$, ${k'_0}^2$, $\sigma_s$, and $\sigma_w$ as a loss function. The formulation of balancing as an optimization problem also makes it possible to incorporate game-specific constraints, for example, requiring Heavy units \textbf{H} to have higher defense than Light ones \textbf{L}, or enforcing that the attack level of Long-Range units \textbf{LR} should be lower than that of Short-Range \textbf{SR} ones.

To obtain credible and reproducible results, thousands of simulations are required for each set of parameters. This raises the computational complexity and makes brute-force search even more impractical. Indeed, to iterate through all possible combinations of unit parameters, we would need to simulate $31^{10} \cdot 10^3 \sim 8.2 \cdot 10^{17}$ games. Even assuming the game engine can simulate $10^4$ games per second, it would have taken approximately $2.6\cdot10^6$ years to complete the search.

To overcome these computational challenges and solve the optimization problem, we have utilized the Covariance Matrix Adaptation Evolution Strategy (CMA-ES) \cite{cma_es}. For each candidate CMA-ES solution, we have simulated 10,000 battles. Since the CMA-ES algorithm is designed for optimization in continuous spaces, we treat our parameters as continuous variables during optimization and round them to the nearest integer when finally evaluating the loss function. This approach allows us to leverage the strengths of CMA-ES while still adhering to the discrete nature of our problem.

\subsection{Analysis of balance parameters}
Depending on the chosen loss function, the optimization yields different sets of attack and defense parameters of the units (see Table~\ref{tab_results}). Although the final $A$ and $D$ values vary across different loss functions, several consistent trends are evident. Most optimization results suggest that the best balance is achieved when the attack levels of \textbf{F} and \textbf{SR} units are the highest, while that of \textbf{LR} unit is the lowest. On the other hand, \textbf{SR} should have relatively low and \textbf{H} relatively high defense levels. Indeed, such a combination of parameters is determined by the units' tactical properties, $AR$ and $MP$.

\begin{table}[h]
\centering
\begin{tabular}{c|cccccc}
$l_{ab}$ & $k^2$ & $k_0^2$ & ${k^\prime}^2$ & ${k_0^\prime}^2$ & $\sigma_s$ & $\sigma_w$ \\
\hline
$k^2$ & \heat{0} & \heat{0.648} & \heat{1.268} & \heat{1.273} & \heat{1.533} & \heat{0.971} \\
$k_0^2$ & \heat{0.648} & \heat{0} & \heat{1.163} & \heat{1.129} & \heat{1.498} & \heat{1.047} \\
${k^\prime}^2$ & \heat{1.268} & \heat{1.163} & \heat{0} & \heat{0.641} & \heat{1.322} & \heat{1.092} \\
${k_0^\prime}^2$ & \heat{1.273} & \heat{1.129} & \heat{0.641} & \heat{0} & \heat{0.936} & \heat{0.711} \\
$\sigma_s$ & \heat{1.533} & \heat{1.498} & \heat{1.322} & \heat{0.936} & \heat{0} & \heat{1.034} \\
$\sigma_w$ & \heat{0.971} & \heat{1.047} & \heat{1.092} & \heat{0.711} & \heat{1.034} & \heat{0} \\
\end{tabular}
\caption{Matrix of the Euclidean distances $l_{ab}$ between the unit parameters optimized using different loss functions.}
\label{tab:similarity_matrix}
\end{table}

To understand how correlated (or independent) the optimal unit parameters found with different loss functions are, we show in Table~\ref{tab:similarity_matrix} the pairwise Euclidean distances
$$l_{ab} = \frac{1}{30} \sqrt{\sum_{i}\left[ (A_{i,a} - A_{i,b})^2 + (D_{i,a} - D_{i,b})^2\right]}$$
between the values of the unit parameters $A$ and $D$ found for different types $i$ of units due to optimization with different loss functions $a$ and $b$. Notice that, by their construction, values $l_{ab}$ are confined in the region $0 \leq l_{ab} \leq \sqrt{10}\approx 3.162$.

It is no surprise that optimizations based on $k^2$ and $k_0^2$, as well as on ${k'}^2$ and ${k_0'}^2$ metrics, give similar results. However, it is interesting that the balancing based on the victory impact $\sigma_w$ leads to results that are also quite similar to those obtained due to pairwise comparisons in semi-random team settings ${k_0'}^2$. Notice that this type of similarity is also observed for different damage functions $DM$ and different sets of tactical parameters $AR$ and $MP$.

Another finding is that the survival rate $s_i$ cannot be balanced for all game units simultaneously. This is infeasible and contradicts the results of other balancing approaches (see Table~\ref{tab_results}). From the other perspective, the set of parameters obtained by optimizing $\sigma_s$ is more distant from the other sets. The faster a unit is, the more quickly it tends to be defeated. Therefore, the optimal tactic for such units is to quickly reach the enemy and engage them, thus allowing Long-Range units more time to act. The latter are more likely to survive because they can attack from a distance. Thus, if players are engaged in many consecutive battles and wish to maximize their units' survival, they should prefer longer-range units.

A strict balance between kills and deaths (parameter $k_0^2$) does not appear to be feasible either, since its minimal value is found to be equal to $0.462$ (see Table~\ref{tab_results}), which is significantly higher than those of other similar metrics. On the other hand, balancing with semi-random teams appears achievable and leads to consistent values across all other defined metrics. In particular, it results in relatively fair victory impact values $w_i$. In contrast, balancing by victory impact $\sigma_w$ also yields small values of ${k'}^2$ and ${k_0'}^2$, making these metrics good candidates for a loss function in the balance optimization problem.

The MSE $\sigma_n$ of the number of units remaining at the end of a game is not a suitable candidate for a loss function, but it can serve as an indicator of balance: if it is too high, the game is likely to be imbalanced. The same applies to the MSE $\varepsilon_a$ of the total number of actions.

Hence, to build a fair and interesting game, one should thoroughly define the damage function and the units' tactical parameters, and then optimize the units' attack and defense parameters using either semi-random team play or victory impact. However, the latter is more unambiguous and does not require assumptions about the composition of semi-random teams, making the MSE $\sigma_w$ of victory impact a more preferable loss function for the optimization problem of unit parameter balancing.

\section{Conclusion}
\label{conclusion}
In the present paper, we introduce a new turn-based strategy game, TBCA, that resembles such classic turn-based strategy games as \textit{Heroes of Might and Magic}. The game has numerous tunable parameters, allowing for many different game configurations and exploration of various game mechanics and strategies. We have derived an expression for the game's state-space complexity and considered a special case of that game with 5 unit types and 8 units per team, showing that the appropriate state-space complexity is about $10^{60}$, which is higher than that of chess.

We have designed a high-performance framework for simulating battles in TBCA, achieving a high game rate, efficient data storage, and transparent integration with AI players. This system includes an in-memory stage for accumulating battle statistics and a real-time WebSocket API for communication with AI players. We have also developed a simple rule-based combat agent that exhibits diverse behaviors across different unit types.

Several metrics for measuring game balance are suggested and analyzed, including pairwise win-loss matrices for same-unit teams and semi-random teams, as well as battle- and unit-specific metrics. The problem of game balance is reduced to balancing unit parameters. The latter is formulated as an optimization problem with 10 parameters, and the CMA-ES algorithm is applied to find optimal unit parameters across different loss functions. Our results suggest that balancing based on victory impact or semi-random team play leads to consistent and fair unit parameters.



\begin{thebibliography}{9}

\bibitem{chinook1}
Schaeffer, J., Culberson, J., Treloar, N., Knight, B., Lu, P., and Szafron, D., 1992. A world championship caliber checkers program. Artificial Intelligence, 53(2--3), 273--289.

\bibitem{chess-champion}
Campbell, M., Hoane Jr, A.J., and Hsu, F.H., 2002. Deep Blue. Artificial Intelligence, 134(1--2), 57--83.

\bibitem{logitello}
Buro, M., 1997. The Othello match of the year: Takeshi Murakami vs. Logistello: Princeton NJ, USA August 4--7, 1997. ICGA Journal, 20(3), 189--193.

\bibitem{arimaa}
Wu, D. J., 2015. Designing a winning Arimaa program. ICGA Journal, 38.1, 19--40.

\bibitem{checkers}
Schaeffer, J., Burch, N., Bjornsson, Y., Kishimoto, A., Muller, M., Lake, R., Lu, P., and Sutphen, S., 2007. Checkers is solved. Science, 317(5844), 1518--1522.

\bibitem{shannon-chess}
Shannon, C.E., 1950. XXII. Programming a computer for playing chess. The London, Edinburgh, and Dublin Philosophical Magazine and Journal of Science, 41(314), 256--275.

\bibitem{arimaa_thesis}
Cox, C.J., 2006. Analysis and implementation of the Game Arimaa. M. Sc. Diss., Universiteit Maastricht, The Netherlands.

\bibitem{go-complexity}
Tromp, J. and Farnebäck, G., 2006. Combinatorics of Go. In International Conference on Computers and Games (pp. 84--99). Berlin, Heidelberg: Springer Berlin Heidelberg.

\bibitem{alpha-go}
Silver, D., Huang, A., Maddison, C.J., Guez, A., Sifre, L., Van Den Driessche, G., Schrittwieser, J., Antonoglou, I., Panneershelvam, V., Lanctot, M., and Dieleman, S., 2016. Mastering the game of Go with deep neural networks and tree search. Nature, 529(7587), 484--489.

\bibitem{alpha-zero}
Silver, D., Hubert, T., Schrittwieser, J., Antonoglou, I., Lai, M., Guez, A., Lanctot, M., Sifre, L., Kumaran, D., Graepel, T., and Lillicrap, T., 2018. A general reinforcement learning algorithm that masters chess, shogi, and Go through self-play. Science, 362(6419), 1140--1144.

\bibitem{stockfish_github}
\url{https://github.com/official-stockfish/stockfish}

\bibitem{stockfish_nn}
\url{https://github.com/official-stockfish/Stockfish/commit/af110e0}

\bibitem{stockfish_winner}
\url{https://tcec-chess.com/#div=sf&game=1&season=28}

\bibitem{starcraft1}
Vinyals, O., Babuschkin, I., Czarnecki, W.M., Mathieu, M., Dudzik, A., Chung, J., Choi, D.H., Powell, R., Ewalds, T., Georgiev, P., and Oh, J., 2019. Grandmaster level in StarCraft II using multiagent reinforcement learning. Nature, 575(7782), 350--354.

\bibitem{minecraft1}
Hafner, D., Pasukonis, J., Ba, J., and Lillicrap, T., 2025. Mastering diverse control tasks through world models. Nature, 640(8059), 647--653.

\bibitem{dota2_1}
Berner, C., Brockman, G., Chan, B., Cheung, V., Dębiak, P., Dennison, C., Farhi, D., Fischer, Q., Hashme, S., Hesse, C., and Józefowicz, R., 2019. Dota 2 with large scale deep reinforcement learning. arXiv Preprint arXiv:1912.06680.

\bibitem{google_football}
Kurach, K., Raichuk, A., Stanczyk, P., Zajac, M., Bachem, O., Espeholt, L., Riquelme, C., Vincent, D., Michalski, M., Bousquet, O., et al. 2020. Google Research Football: a novel reinforcement learning environment. In Proceedings of the AAAI Conference on Artificial Intelligence, Vol. 34. pp. 4501--4510.

\bibitem{chess_new}
Chacoma, A. and Billoni, O.V., 2025. Emergent complexity in the decision-making process of chess players. Sci. Rep., 15(1), 23234.

\bibitem{diversity1}
Dasgupta, R. and Kliem, J., 2023. Improved reinforcement learning in asymmetric real-time strategy games via strategy diversity: a case study for Hunting-of-the-Plark game. International Journal of Serious Games, 10(1), 19--38.

\bibitem{diversity2}
Wong, K.M., Lim, S.W., and Gao, Z., 2005. Effects of diversity on multiagent systems: minority games. Phys. Rev. E, 71(6), 066103.

\bibitem{diversity3}
Perez-Nieves, N., Yang, Y., Slumbers, O., Mguni, D.H., Wen, Y., and Wang, J., 2021, July. Modelling behavioural diversity for learning in open-ended games. In International Conference on Machine Learning (pp. 8514--8524). PMLR.

\bibitem{chess_zipf}
Blasius, B. and Tönjes, R., 2009. Zipf's law in the popularity distribution of chess openings. Phys. Rev. Lett., 103(21), 218701.

\bibitem{chess_memory}
Schaigorodsky, A.L., Perotti, J.I., and Billoni, O.V., 2016. A study of memory effects in a chess database. PLoS One, 11(12), e0168213.

\bibitem{alphazero_zipf}
Neumann, O. and Gros, C., 2026. Alphazero neural scaling and Zipf's law: a tale of board games and power laws. Advances in Neural Information Processing Systems, 38, 58351--58381.

\bibitem{balancing_wargames}
Long, G.E., Perez-Liebana, D., and Samothrakis, S., 2023, August. Balancing wargames through predicting unit point costs. In 2023 IEEE Conference on Games (pp. 1--8). IEEE.

\bibitem{shapley}
Shapley, L.S., 1953. Stochastic games. Proc. Natl. Acad. Sci., 39(10), 1095--1100.

\bibitem{mpe1}
Doraszelski, U. and Escobar, J.F., 2010. A theory of regular Markov perfect equilibria in dynamic stochastic games: genericity, stability, and purification. Theoretical Economics, 5(3), 369--402.

\bibitem{mpe2}
He, W. and Sun, Y., 2017. Stationary Markov perfect equilibria in discounted stochastic games. Journal of Economic Theory, 169, 35--61.

\bibitem{formula}
Attia, L. and Oliu-Barton, M., 2019. A formula for the value of a stochastic game. Proc. Natl. Acad. Sci., 116(52), 26435--26443.

\bibitem{complexity}
Daskalakis, C., Golowich, N., and Zhang, K., 2023, July. The complexity of Markov equilibrium in stochastic games. In The Thirty Sixth Annual Conference on Learning Theory (pp. 4180--4234). PMLR.

\bibitem{tubstap}
Fujiki, T., Ikeda, K., and Viennot, S., 2015. A platform for turn-based strategy games, with a comparison of Monte-Carlo algorithms. In 2015 IEEE Conference on Computational Intelligence and Games (CIG) (pp. 407--414). IEEE.


\bibitem{stratega}
Dockhorn, A., Grueso, J.H., Jeurissen, D., and Liebana, D.P., 2020. STRATEGA: a general strategy games framework. In AIIDE Workshops (Vol. 22, pp. 43--44).

\bibitem{adapta}
Bergsma, M.B.M. and Spronck, P., 2008. Adaptive spatial reasoning for turn-based strategy games. In Proceedings of the AAAI Conference on Artificial Intelligence and Interactive Digital Entertainment (Vol. 4, No. 1, pp. 161--166).

\bibitem{thurn}
Schadd, F.C., 2009. Monte-Carlo search techniques in the modern board game Thurn and Taxis (Doctoral dissertation, MS thesis, Maastricht Univ., Netherlands).

\bibitem{balance1}
Leigh, R., Schonfeld, J., and Louis, S.J., 2008. Using coevolution to understand and validate game balance in continuous games. In Proceedings of the 10th Annual Conference on Genetic and Evolutionary Computation (pp. 1563--1570).

\bibitem{balance2}
Mahlmann, T., Togelius, J., and Yannakakis, G.N., 2012. Evolving card sets towards balancing dominion. In 2012 IEEE Congress on Evolutionary Computation (pp. 1--8). IEEE.

\bibitem{balance3}
Elfeky, E.Z., Elsayed, S., Marsh, L., Essam, D., Cochrane, M., Sims, B., and Sarker, R., 2021. A systematic review of coevolution in real-time strategy games. IEEE Access, 9, 136647--136665.

\bibitem{liitman}
Littman, M.L., 1994. Markov games as a framework for multiagent reinforcement learning. In Machine Learning Proceedings 1994 (pp. 157--163). Morgan Kaufmann.

\bibitem{balance00}
Long, G.E., Perez-Liebana, D., and Samothrakis, S., 2024. STEP: a framework for automated point cost estimation. IEEE Transactions on Games, 16(4), 927--936.

\bibitem{cma_es}
Hansen, N., 2016. The CMA evolution strategy: A tutorial. arXiv Preprint arXiv:1604.00772.


%

\end{thebibliography}
\end{document}